\documentclass[paper]{JHEP3}
\usepackage[centertags]{amsmath}
\usepackage{amsfonts} \usepackage{amssymb} \usepackage{amsthm}
\usepackage{graphicx}
\usepackage{psfrag}
\newcommand{\bra}[1]{\langle{#1}|}
\newcommand{\ket}[1]{|{#1}\rangle}
\def\one{{\rm 1\kern -.9mm l}}                             %
\newcommand{\braket}[2]{\langle{#1}|{#2}\rangle}
\def\beq{\begin{equation}}
\def\eeq{\end{equation}}
\def\beq{\begin{equation}}
\def\eeq{\end{equation}}
\def\beqa{\begin{eqnarray}}
\def\eeqa{\end{eqnarray}}
\newcommand{\eq}[1]{eq. (\ref{#1})}
\newcommand{\Tr}{\mathrm{Tr}\,}
\def\gh{{\rm gh}}
\def\ii{\mathrm{i}}
\def\ee{\mathrm{e}}                                                           
\newcommand{\Z}{\mathbb{Z}}

\newcommand{\comm}[2]{\left[ #1,#2\right]}

\title{Polyakov loop correlators from D0-brane interactions
in bosonic string theory%
\thanks{Work partially supported by the European Community's Human Potential
Programme under contract MRTN-CT-2004-005104 ``{Constituents, Fundamental
Forces and Symmetries of the Universe}'', by the European Commission TMR
programme HPRN-CT-2002-00325 (EUCLID) and by the Italian M.I.U.R under contract
PRIN-2003023852 ``{Physics of fundamental interactions: gauge theories, gravity
and strings}''.}}
\author{M. Bill\'o, M. Caselle
\\
Dipartimento di Fisica Teorica, Universit\`a di Torino\\
and Istituto Nazionale di Fisica Nucleare - sezione di Torino \\
Via P. Giuria 1, I-10125 Torino, Italy
}
\abstract{In this paper we re-derive the effective Nambu-Goto theory result for
the Polyakov loop correlator, starting from the free bosonic string and using a
covariant quantization. The boundary conditions are those of an open string
attached to two D0-branes at spatial distance $R$, in a target space with
compact euclidean time. The one-loop free energy contains topologically distinct
sectors corresponding to multiple covers of the cylinder in target space
bordered by the Polyakov loops. The sector that winds once reproduces exactly
the Nambu-Goto partition function. In our approach, the world-sheet duality
between the open and closed channel is most evident and allows for an explicit
interpretation of the free energy in terms of tree level exchange of closed
strings between boundary states. Our treatment is fully consistent only in
$d=26$; extension to generic $d$ may be justified for large $R$, and is
supported by Montecarlo data. At shorter scales, consistency and Montecarlo data
seem to suggest the necessity of taking into account the Liouville mode of
Polyakov's formulation.
}
\preprint{DFTT/13/2005}

\begin{document}

\section{Introduction}
\label{sec:intro}
It is a long-standing belief \cite{early} that the confining regime of
non-Abelian gauge theory should be described by an effective string theory
describing the fluctuations of the color flux tube. Many theoretical insights
and proposals have been put forward, while the development of lattice gauge
theories (LGT) provides a better and better numerical test ground for the various
models.  

One of the main predictions which can be extracted from an effective string
model and then tested in LGT simulations is the potential $V(R)$ between two
external, massive quark and anti-quark sources in a pure glue theory. This
potential can be obtained by considering, for instance, a rectangular Wilson
loop $W(L,R)$ of sides $L$ and $R$, for which $<W(L,R)>\sim \ee^{-L V(R)}$ in
the limit of large $L$. In the confining phase, the area law corresponds to a
linear potential $V(R) = \mathcal{T} R +\ldots$. In a string interpretation, the
area term $\mathcal{T} L R$ in the exponent of the Wilson loop is the classical
action of the string model; $\mathcal{T}$ represents the string
tension\footnote{We denote the string tension with $\mathcal{T}$ rather than
with the usual notation $\sigma$ to avoid confusion with the spatial coordinate
of the string world sheet.}. Upon quantization of the string model, we expect
corrections to this classical potential. 

In a seminal paper, L\"uscher, Symanzik and Weisz~\cite{lsw}, starting from the
loop equations satisfied by the Wilson loops, derived the leading correction for
large $R$. They found
\begin{equation}
\label{lterm}
V(R)= \mathcal{T}\, R - \frac{\pi}{24}\frac{d-2}{R} +
O\left(\frac{1}{R^2}\right)~.
\end{equation} 
Their computation, and subsequent Ref. \cite{l81}, linked this correction to the
universal quantum contribution of $d-2$ massless modes corresponding to the
transverse fluctuations of the string joining the quark--anti-quark pair. In
this spirit,  most of the theoretical calculations and of the comparisons with
the lattice results in these last years were performed in an \emph{effective}
description via the $c=d-2$ two-dimensional conformal field theory of free
bosons defined over the space-time surface of interest and with the appropriate
boundary conditions; for instance, over the rectangle bordered by the Wilson
loop, with Dirichlet boundary conditions. The picture may be refined by allowing
a set of higher order interactions among the fields of the theory, whose precise
form distinguishes the various effective theories. In a string picture, the
underlying string model should determine a specific form of the effective
theory, and an expression of the potential that extends \eq{lterm} to finite
values of $R$.

An observable which presents a lot of interest in this respect is the correlator
of two Polyakov loops at spatial distance $R$ in the gauge theory at finite
temperature $1/L$. On the one hand, this quantity can be measured with very high
precision on the lattice. On the other hand, in a string
interpretation, the correlation is due to the cylindric world-sheet spanned by
the stretched string and is therefore associated to the partition function of
the effective string model, and not just to its ground state energy $V(R)$. The
Polyakov loop is thus very useful to discriminate between different models. 

The simplest and most natural string model is the Nambu-Goto \cite{nambu-goto}
one. For the Nambu-Goto string with boundary conditions corresponding to fixed
ends in the spatial directions (the static quark and anti-quark) Alvarez
\cite{alvarez81} (for $d\to\infty$) and Arvis \cite{Arvis:1983fp}, with a formal
quantization, obtained the energy spectrum 
\beq
E_n(R)=\mathcal{T} R \sqrt{1+\frac{2\pi}{\mathcal{T} R^2}(n-\frac{d-2}{24})}~.
\label{en}
\eeq
The static potential equals the lowest energy level: $V(R) = E_0(R)$,
reproducing \eq{lterm} for large $R$. The partition function is thus
\beq
\label{zng}
Z=\sum_n w_n e^{-L E_n(R)}~,
\eeq
$w_n$ being the usual multiplicities of the  bosonic string. The derivation of
eq.s (\ref{en}) and (\ref{zng}) in \cite{Arvis:1983fp} uses the
re-parametrization invariance of the world-sheet to reach the conformal gauge
(where the Nambu-Goto action is equivalent to the free string action) and the
residual conformal invariance to fix a light-cone type gauge (which is sometimes
denoted as ``physical gauge''). This leaves as only independent dynamical
variables the transverse modes, which become oscillators upon quantization. The
energy, in particular, is re-expressed in terms of the occupation number $n$ of
these oscillators and the spectrum \eq{en}, as remarked in \cite{Olesen:1985pv},
stems from the analogue of the standard mass formula $M^2 = \mathcal{T}[n -
(d-2)/24]$ for the bosonic open string with free ends.  

This bosonic string model, of course, is truly consistent at the quantum level
only if $d=26$: as usual in light-cone type gauges, Lorentz invariance is
otherwise broken. It was however noticed in \cite{Olesen:1985pv} that the
coefficient of the anomaly vanishes for $R\to\infty$, so that in this regime the
model could be consistent. 

In these last years, thanks to  various remarkable improvements in lattice simulations
~\cite{lw01,cfghp97,fep00,chp03} the effective string picture could be tested with a 
very high degree of precision and confidence
\cite{chp03}-\cite{mt04}. The picture which emerges is that at large inter-quark
distances and low temperatures the Nambu-Goto effective string \eq{en} correctly
describes the Montecarlo data. Moreover, this result is
universal, meaning that it does not depend on the particular gauge group under
study (the same behaviour is observed in models as different as the
$\mathbb{Z}_2$ gauge model in $(2+1)$ dimensions \cite{chp05} and the
$\mathrm{SU}(3)$ LGT in $(3+1)$ dimensions \cite{lw02}). As the
inter-quark distance decreases and/or the temperature increases (i.e. as the
de-confinement transition is approached) clear deviations from this picture are
observed and the universality mentioned above is partially
lost~\cite{chp04,chp05,jkm03,jkm04}.

These deviations could well be connected to the inconsistency of the model at
$d<26$ becoming more and more relevant as $R$ decreases. A consistent quantum
formulation in $d<26$ can be sought in the Polyakov formulation
\cite{Polyakov:1981rd}. We will briefly comment on this possibility in sec.
(\ref{sec:conclusions}).

In this paper, we re-derive the effective Nambu-Goto theory, and in particular
the result \eq{zng} for the Polyakov loop correlator, starting from the free
bosonic string and using a covariant quantization%
\footnote{In a covariant quantization, the conformal invariance of the free
string model is not gauge-fixed by identifying the world-sheet coordinates with
some directions in the target space. The constraints corresponding to the
conformal symmetry (Virasoro constraints) are imposed at the quantum level
\emph{\'a la} Gupta-Bleuer (old covariant quantization) or via BRST
quantization. See, for instance, \cite{gsw} or \cite{polbook} for a review; some
more details are recalled here in sec. \ref{sec:PolyaD0}.}.
The boundary conditions (the same as in Arvis' paper) are described in modern
terms as those of an open string attached to two D0-branes at spatial distance
$R$. We work at finite temperature, i.e., in a target space with compact
euclidean time, and compute the free energy for such open strings. This is
nothing but the well-known Polchinski derivation of the interaction between two
D-branes \cite{pol_comp}, adapted to the present case. We do not impose any
light-cone-like or physical gauge, so we keep the string world-sheet distinct
from the target-space surface bordered by the Polyakov loops. In fact, the free
energy contains different topological sectors corresponding to multiple covers
of the cylinder in target space bordered by the Polyakov loops. The sector that
winds once reproduces exactly \eq{zng}. Thus, the expression of the Polyakov
loop correlator and the inter-quark potential obtained from the covariant
quantization of the free bosonic string is the one of the Nambu-Goto effective
string, and not the one%
\footnote{In this case, the energy spectrum corresponds to the second-order
truncation of the square-root in \eq{en}, yielding a partition function which is
simply $Z_{(0)} = \ee^{\mathcal{T}LR}\eta(\ii\, \frac{L}{2R})$,  $\eta$ being
the well known Dedekind function.} 
obtained from the effective free bosonic string, i.e. the CFT of $d-2$ free
bosons which is the simplest element in the universality class of \eq{lterm}. 

Just as for Arvis, our treatment is fully consistent only in $d=26$; extension
to generic $d$ may be justified for large $R$, and is supported by Montecarlo
data as already described above. 

We think that the re-derivation of the Nambu-Goto effective theory presented
here may have some advantages. It dwells on a standard computation in the
framework of the free bosonic string, the open string free energy at one loop,
for which simple operatorial methods are effective. Not having fixed a
``physical gauge'', the world-sheet duality between the open and closed channel
is most evident and allows for an explicit interpretation of the free energy in
terms of tree level exchange of closed string states between boundary states.
Our formulation is well suited in principle to study the contributions to the
inter-quark potential from string interactions, which in our language would mean
wrapping the Polyakov string on surfaces (bordered by the Polyakov loops) with
handles, as well as to investigate different observables such as Wilson loops or
interfaces. It could also be interesting to investigate the possible relevance
in the gauge theory of the contributions to the free energy with different
winding numbers. Finally, since the covariant treatment basically coincides with
Polyakov's formulation upon neglecting the Liouville field, one may try to
correct its results at finite scale $R$ by including the effect of the Liouville
theory, see the brief discussion in sec. \ref{sec:conclusions}. 

This paper is organized as follows. Sec. \ref{sec:PolyaD0} contains the
main computation. In particular, in subsec. \ref{subsec:osD0} we compute the
open string free energy, in subsec. \ref{subsec:mod_transf} we perform its
modular transformation and in subsec. \ref{subsec:cl_bs} we give a detailed
re-interpretation of the modular transformed expression in the closed string
channel. In sec. \ref{sec:conclusions} some conclusions and speculations are
presented.

\section{Polyakov loop correlators and strings between D0-branes}
\label{sec:PolyaD0}
The Polyakov loop is the trace of the temporal Wilson line induced by the
presence of a static quark minimally coupled to the non-abelian gauge field. In
a stringy perspective, the quark represents the end-point of a string. The
v.e.v. of $P(\vec R)$ is related to the free energy of this static quark:
$\langle P(\vec R) \rangle = \ee^{-F}$, and a non-zero value for this
v.e.v. signals de-confinement, as having an isolated quark requires a finite
energy. 
\FIGURE{
\psfrag{T}{\small $T$}
\psfrag{x0}{\small $x^0$}
\psfrag{xvec}{\small $\vec x$}
\psfrag{rvec}{\small $\vec R$}
\psfrag{D0}{\small D$0$}
\psfrag{q}{\small $q$}
\psfrag{qbar}{\small $\bar q$}
\includegraphics[width=5.5cm]{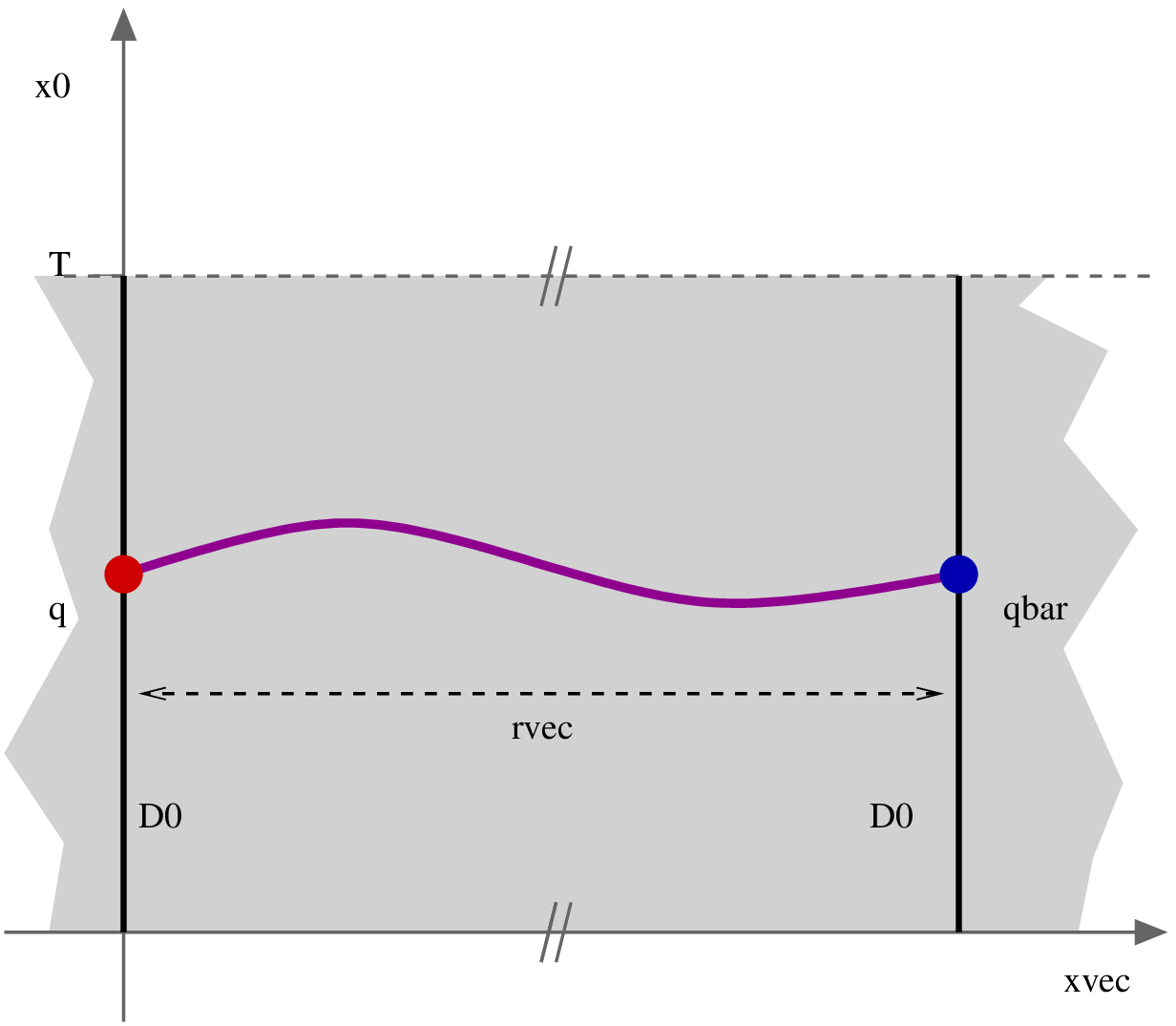}%
\caption{\label{fig:setup} 
\small Let us consider two Polyakov loops at spatial distance $\vec
R$. An open string connecting the two external static quarks obeys boundary
conditions corresponding to D$0$-branes on a compact space.}
}
The Polyakov loop is the order parameter of the $\Z_N$ global symmetry which
appears when the $\mathrm{SU}(N)$ gauge theory is regularized on a lattice with
a finite-temperature geometry (i.e. periodic boundary conditions in the ``time''
direction). This additional symmetry coincides with the center of the original
gauge group and is broken in the de-confined phase.

An observable which can be measured with great accuracy on the lattice is the
connected correlator of two spatially separated Polyakov loops, 
\beq
\label{corr_pol}
\langle P(\vec 0) P(\vec R)\rangle_{\mathrm c}~.
\eeq
In the string picture, see Fig. \ref{fig:setup}, the correlation is due to the
strings connecting the two external, static quark and anti-quark that span the
two Polyakov loops.  We take the point of view that such strings can be
described by the standard bosonic string theory in the $d$-dimensional
space-time under consideration, which is flat, but with compact Euclidean
time
\beq
\label{comp_time}
x_0\sim x_0 + L~.
\eeq 
Using the first order formulation, the string action in the conformal gauge
reads simply
\beq
\label{sac}
S = \frac{1}{4\pi\alpha'}\int d\tau \int_0^\pi d\sigma\,
\left[(\partial_\tau X^M)^2 + (\partial_\sigma X^M)^2\right] +
S_{\mathrm{gh.}}~,
\eeq
where $\sigma\in [0,\pi]$ parametrizes the spatial extension of the string and
$\tau$ its proper time evolution. The string tension $\mathcal{T}$ is given in
this notation by
\beq
\label{sten}
\mathcal{T} = \frac{1}{2\pi\alpha'}~.
\eeq 
The fields $X^M(\tau,\sigma)$, with $M=0,\ldots,d-1$, describe the embedding of
the string world-sheet in the target space and form the 2-dimensional CFT of $d$
free bosons. The term $S_{\mathrm{gh.}}$ in \eq{sac} is the action for the ghost
and anti-ghost fields (traditionally called $c$ and $b$) that arise from the
Jacobian for fixing the conformal gauge. We do not really need here its explicit
expression, see \cite{gsw} or \cite{polbook} for reviews. In the conformal gauge
the world-sheet metric is of the form $g_{\alpha\beta} = \ee^\phi
\delta_{\alpha\beta}$, and corresponds to a CFT of central charge
$c_{\mathrm{gh.}} = -26$. The scale factor $\ee^\phi$ decouples at the classical
level, but this property persists at the quantum level only if the anomaly
parametrized by the total central charge $c= d - 26$ vanishes. We will
nevertheless proceed in the case of general $d$, according to the discussion in
the introduction.
 
The open string joining the $q$-$\bar q$ static quarks as in Fig.
\ref{fig:setup} obeys Neumann boundary conditions at both ends in the time
direction:
\beq
\label{nbcp}
\left.\partial_\sigma X^0(\tau,\sigma)\right|_{\sigma=0,\pi} = 0~,
\eeq
while it satisfies Dirichlet boundary conditions in the spatial directions:
\beq
\label{dbcp}
\vec X(\tau,0) = 0~,
\hskip 0.8cm
\vec X(\tau,\pi) = \vec r~.
\eeq
These conditions constrain the endpoints to two lines (1-dimensional ``branes'')
which are nowadays known as D$0$-branes. The mode expansion of the $X^M$ fields
with such boundary conditions is
\beq
\label{mode_op}
\begin{aligned}
X^0(\tau,\sigma) & = \hat x^0 + 2\alpha' \hat p^0\tau + \ii 
\sqrt{2\alpha'} \sum_{n\not = 0} \frac{\alpha^0}{n}\ee^{-\ii n \tau}
\cos n\sigma~,
\\
\vec X(\tau,\sigma) & = \frac{\vec R}{\pi} \sigma - 
\sqrt{2\alpha'} \sum_{n\not = 0} \frac{\vec\alpha}{n}\ee^{-\ii n \tau}
\sin n\sigma~,
\end{aligned}
\eeq
Upon canonical quantization, the oscillators $\alpha^M_n$ obey the algebra
\beq
\label{osc_open}
\comm{\alpha^M_m}{\alpha^N_n} = m\, \delta_{m+n,0}\,\delta^{MN}~.
\eeq
The eigenvalues $p^0$ of the momentum operator $\hat p^0$ are discrete, because
of the periodicity \eq{comp_time}:
\beq
\label{eigen_p0}
\hat p^0 \to \frac{2\pi n}{L}~.
\eeq
The generators of the conformal transformations are called Virasoro generators
and traditionally indicated as $L_m$. In particular, $L_0$ generates the
world-sheet dilations and corresponds to the Hamiltonian derived from the action
\eq{sac}. It receives therefore contributions from the bosons and the ghost
system: $L_0= L_0^{(X)} + L_0^{(\mathrm{gh.})}$, and we have
\beq
\label{L0Xo}
L_0^{(X)} = \alpha' (\hat p^0)^2 + 
\frac{R^2}{4\pi^2 \alpha'} +
\sum_{n=1}^\infty N_n - \frac{d}{24}~,
\eeq
where $N_n = \sum_M\alpha^M_{-n}\cdot \alpha^M_{n}$ is the occupation number for
the oscillators $\alpha^M_n$, and $d/24$ is the ($\zeta$ function regularized%
\footnote{In Ref. \cite{Orland:2001rq} some interesting words of caution were
raised regarding the use of $\zeta$-function regularization for the string with
the present boundary conditions.})
normal ordering constant. The contribution $R^2/(4\pi^2\alpha')$ represents the
energy needed to stretch the string between the two branes. For the $b,c$ ghost
system we have, see for instance \cite{gsw},
\beq
\label{L0gh}
L_0^{\gh} = \mbox{non-zero modes} + \frac{1}{12}~. 
\eeq
The $b,c,$ are indeed anti-commuting bosonic fields, and they get a normal
ordering constant of the opposite sign with respect to the $X$'s and, in the
trace of \eq{fe1} they will contribute exactly the inverse of the non-zero mode
part of two bosonic directions. 

\subsection{Open string free energy}
\label{subsec:osD0}
Let us now compute the one-loop free energy of the (non-critical) open strings
with their endpoints attached to the two different D0-branes. The expression to
be considered is%
\footnote{We consider a given orientation of the string.} 
\beq
\label{fe1}
\mathcal{F} =  L\,\int_0^\infty\frac{dt}{2t}\,
\Tr q^{L_0}~.
\eeq
In this expression, we integrate over the single real modulus $t$ of the
world-sheet surface, which is a cylinder, as we have to do in our first-order
formulation. The factor $L$ represents the volume of the only target space
direction along which the excitations can propagate, namely the Euclidean time.
In \eq{fe1} we introduced 
\beq
\label{defq}
q = \exp(-2\pi t)
\eeq
The trace in \eq{fe1} decomposes in a trace over non-zero modes and a zero-mode
part. For the non-zero modes, including also the normal-ordering constant, we
get for each bosonic direction the usual result
\beq
\label{etatrace}
q^{-\frac{1}{24}}\prod_{r=1}^\infty \frac{1}{1 - q^r} = \frac{1}{\eta(\ii t)}~.
\eeq

The only operatorial zero mode is the momentum $\hat p^0$ appearing in the $X^0$
field; the distance $\vec r$ between the branes appears instead as a numerical
parameter in the expansion of the $\vec X$ fields. Since the $0$-th direction is
compactified according to \eq{comp_time}, the eigenvalues of $p^0$ are
quantized, see \eq{eigen_p0}: $p^0 = 2\pi n/L$. The corresponding trace, which
in the non-compact case is given by an integral, requires therefore the discrete
sum $(1/L)\sum_n$.

Taking into account also the zero-mode contributions to $L_0$ and the relation,
\eq{defq}, between $q$ and $t$ the free energy \eq{fe1} is expressed as
\beq
\label{fe2}
\mathcal{F} = \int_0^\infty\frac{dt}{2t}\,
\sum_{n=-\infty}^\infty
\ee^{-2\pi t\left(\frac{4\pi^2\alpha'\, n^2}{L^2} +
\frac{R^2}{4\pi^2\alpha'}\right)} 
\left(\frac{1}{\eta(\ii t)}\right)^{d-2}~,
\eeq
where the exponent of $d-2$ for the non-zero mode trace is due to the fact that
the ghost contribution cancels exactly the non-zero modes of two bosonic
directions. In the effective interpretation, these two coordinates are the time
one, $X^0$, and one of the spatial ones; as a result, $d-2$ spatial transverse
coordinates are left.

We can now Poisson re-sum over the integer $n$ labelling the momentum: 
\beq
\label{poisson}
\sum_{n=-\infty}^\infty \exp\left(-\frac{8\pi^3 \alpha'\, t}{L^2} n^2\right)
= \sqrt{\frac{L^2}{8\pi^2\alpha'\, t}}
\sum_{m=-\infty}^\infty \exp\left(-\frac{L^2}{8\pi\alpha'\, t} m^2\right)~.
\eeq
In this dual expansion, the integer $m$ labels the topologically distinct
sectors in which the string world-sheet winds $m$ times around the compact time
direction. Notice that winding in one direction or the opposite yields the same
contribution, as only $m^2$ occurs. We can thus write
\beq
\label{mexp}
\mathcal{F} = \mathcal{F}^{(0)} + 2 \sum_{m=1}^\infty \mathcal{F}^{(m)}~,
\eeq
with 
\beq
\label{fe3}
\mathcal{F}^{(m)} = \frac{L}{\sqrt{8\pi^2\alpha'}}\int_0^\infty
\frac{dt}{2t^{\frac{3}{2}}}
\ee^{-\frac{L^2\, m^2}{8\pi\alpha'\, t} -t \frac{R^2}{2\pi\alpha'}}
\left(\frac{1}{\eta(\ii t)}\right)^{d-2}~.
\eeq
In the case $m= \pm 1$, the string end-points trace out in the target space the
two Polyakov loops, and the string world-sheet has exactly the topology of the
cylinder bordered by the Polyakov loop whose fluctuations are assumed to be
described by the ``effective'' Nambu-Goto theory in the usual treatment. Let us
see how our computation, in this topological subsector, is related to such a
description.

First of all, in \eq{fe3} we expand in series of $q$ the infinite products in
eq.s (\ref{etatrace},\ref{fe2}):
\beq
\label{etaexp}
\left(\prod_{r=1}^\infty \frac{1}{1 - q^r}\right)^{d-2}
= \sum_{k=0}^\infty w_k q^k
\eeq
(for $d=3$ we have simply $w_k=p_k$, the number of partitions of the integer
$k$). Having done this, the integration over $t$ in \eq{fe3} can be performed%
\footnote{In the case $m\not= 0$ we use the integral
\beq
\label{int32}
\int_0^\infty \frac{dt}{t^{\frac{3}{2}}}\,\ee^{-\frac{\alpha^2}{t}- \beta^2 t}
= \frac{\sqrt{\pi}}{|\alpha|} \ee^{-2|\alpha|\, |\beta|}
\eeq
with
$\alpha^2 = L^2 m^2/(8\pi\alpha')$ and
$\beta^2 = \frac{R^2}{2\pi\alpha'} + 2\pi \left(k - \frac{d-2}{24}\right)$.}
obtaining, in the case $m\not=0$, 
\beq
\label{Fmr}
\mathcal{F}^{(m)} = \frac{1}{2|m|} \sum_k w_k \ee^{-|m| L E_k(R)}~,
\hskip 0.8cm
(m\not= 0)~,
\eeq
where 
\beq
\label{Zs2}
E_k(r) = \frac{R}{4\pi\alpha'}\sqrt{1 + \frac{4\pi^2\alpha'}{R^2}
\left(k - \frac{d-2}{24}\right)}
\eeq
are nothing else but the Nambu-Goto energy levels of Alvarez and Arvis, 
see \eq{en}. In particular, from the $m=\pm 1$ cases we get 
\beq
\label{mpm1}
2\mathcal{F}^{(1)} = Z(R)~,
\eeq
where $Z(R)$ is the Nambu-Goto partition function of \eq{zng}.

The case $m=0$ corresponds exactly (apart from the volume factor $L$ being
finite) to the usual result one gets in the non-compact situation:
\beq
\label{znc}
\mathcal{F}^{(0)} = L \int_0^\infty \frac{dt}{2t}
\frac{1}{\sqrt{8\pi^2\alpha't}}\left(\frac{1}{\eta(\ii
t)}\right)^{d-2}\,\ee^{-\frac{t R}{2\pi\alpha'}}~, 
\eeq
see for instance \cite{Polchinski:1996na}. Using the expansion \eq{etaexp}, the
integration over $t$ can be easily carried out, with the result
\beq
\label{Zm0}
\mathcal{F}^{(0)} =  - L \sum_k w_k E_k(R)~.
\eeq

\subsection{Modular transformation to the closed channel}
\label{subsec:mod_transf}
Changing integration variable to $s =1/t$ and taking advantage of the modular
properties of the Dedekind eta function:
\beq
\label{etamod}
\eta(\ii/s) = s^{\frac 12}\, \eta(\ii s)
\eeq
the expression \eq{fe3} of the free energy in the $m$-th sector can be written
as
\beq
\label{mod3}
\begin{aligned}
\mathcal{F}^{(m)} &= \frac{L}{\sqrt{8\pi^2\alpha'}}\int_0^\infty
\frac{ds}{2s} s^{\frac {3-d}{2}} 
\ee^{-\frac{L^2\, m^2}{8\pi\alpha'} s - \frac{R^2}{2\pi\alpha'\, s}}
\left(\frac{1}{\eta(\ii s)}\right)^{d-2}
\\
& = \frac{L}{\sqrt{8\pi^2\alpha'}}
\sum_k w_k\, 
\int_0^\infty \frac{ds}{2s} s^{\frac {3-d}{2}}
\ee^{-\left[\frac{L^2\, m^2}{8\pi\alpha'} + 
2\pi\left(k - \frac{d-2}{24}\right)\right]s -
\frac{R^2}{2\pi\alpha'\, s}}~.
\end{aligned} 
\eeq
In terms of the variable $z = \pi\alpha' s/2$ this becomes
\beq
\label{mod4}
\mathcal{F}^{(m)} = \frac{L}{\sqrt{8\pi^2\alpha'}}
\left(\frac{2}{\pi\alpha'}\right)^{\frac{3-d}{2}}
\sum_k w_k\, 
\int_0^\infty \frac{dz}{2z} z^{\frac {3-d}{2}}
\ee^{-M^2(m,k)\, z -
\frac{R^2}{4z}}~,
\eeq
with
\beq
\label{cl_mass}
\begin{aligned}
M^2(m,k) &= \frac{4}{\alpha'}\left(k - \frac{d-2}{24}\right) +
\left(\frac{m L}{2\pi\alpha'}\right)^2
\\
& = (m\mathcal{T} L)^2\left[1+\frac{8\pi}{\mathcal{T} L^2 m^2}
\left(k - \frac{d-2}{24}\right)\right]~.
\end{aligned}
\eeq
In the second line above we wrote the expression in such a way that it can be
easily compared, for $m=1$, with eq. (C5) of \cite{lw04}, see the discussion at
the end of this section.
 
The integral appearing in \eq{mod4} is proportional to the propagator of a
scalar field of mass $M^2$ over the distance $\vec R$ between the two D0-branes
along the $d-1$ spatial directions. Indeed, such a propagator is given by
\beq
\label{scal_prop}
\begin{aligned}
G(R;M) & = \int \frac{d^{d-1}p}{(2\pi)^{d-1}} 
\frac{\ee^{\ii \vec p\cdot\vec R}}{p^2+ M^2} =
\int_0^\infty dz  \int \frac{d^{d-1}p}{(2\pi)^{d-1}} \ee^{- z (p^2 + M^2) + \ii
\vec p\cdot\vec r}
\\
& = \frac{1}{(4\pi)^{\frac{d-1}{2}}} \int_0^\infty \frac{dz}{z}
z^{\frac{3-d}{2}} \ee^{-M^2 z - \frac{R^2}{4z}}
= \frac{1}{2\pi} \left(\frac{M}{2\pi R}\right)^{\frac{d-3}{2}}
K_{\frac{d-3}{2}}(MR)~.
\end{aligned}
\eeq
The free energy \eq{mod4} can therefore be seen as a collection of tree-level
exchange diagrams between the D0-branes; the exchanged particles have
squared mass $M^2(k,m)$ given by \eq{cl_mass}. 

This picture nicely agrees with what dimensional reduction and the so called
Svetitsky-Yaffe conjecture~\cite{sy} suggest on the behaviour of the Polyakov
loop correlator as the de-confinement temperature is approached from below.
According to this conjecture, in the vicinity of the de-confinement point, if
the transition is continuous, a $d$-dimensional LGT with gauge group $G$ can be
effectively described by a $d-1$ dimensional spin model with symmetry group the
center of $G$. In this representation the Polyakov loops become the spins of the
underlying model and the Polyakov loop correlator is nothing else than the
standard spin-spin correlator%
\footnote{Notice as a side remark that  the closed string channel discussed here
is the one which better describes the high temperature behaviour of the Polyakov
loop correlators where $L$ is in general much smaller than the interquark
distance $R$.}. 
This correlator will depend on the symmetry group and will be in general very
complicated; however, at large distance it will be dominated by the contribution
of the lowest mass particle in the spectrum, whose contribution to the
correlation function, in $d-1$ dimensions, will be represented exactly by the
Bessel function $K_{\frac{d-3}{2}}(mR)$, with $m\equiv M(1,0)$ being the mass of
this particle. Thus the Nambu-Goto string description is fully compatible with
our qualitative understanding of the high temperature behaviour of the gauge
theory.

Collecting all the numerical factors, we can write
\beq
\label{zasex}
\mathcal{F}^{(m)} = L \frac{T_0^2}{4}
\,\sum_k w_k\, G\left(R;M(m,k)\right)~.
\eeq 
where $T_0$ is the 0-brane tension, in accordance with the standard result for
the tension $T_p$ of D$p$-branes in bosonic string theory, see for instance
\cite{Polchinski:1996na,DiVecchia:1999rh}:
\beq
\label{Tp}
T_p = \sqrt{\frac{\pi}{2^{\frac d2 - 5}}}\left(2\pi\sqrt{\alpha'}\right)^{\frac
d2 - 2 - p}~.
\eeq
which, for $p=0$, gives
\beq
\label{T0}
T_0^2=8\pi\left(\frac{\pi}{\mathcal{T}}\right)^{\frac{d}{2}-2}
\eeq
The exchanged states are \emph{closed string states}, with $k$ representing the
total oscillator number, and $m$ the wrapping number of the string around the
compact time direction. 

The modular transformation to the closed channel of the case $m=1$, i.e., for
the NG partition function \eq{zng}, was performed in \cite{lw04}; let us compare
our findings to that reference. The mass $M$ of the fields exchanged by the
$D_0$ branes, see eq. (\ref{cl_mass}),  coincide for $m=1$ with the ``closed
string energies'' reported in eq. (C5) of \cite{lw04}. With this identification
eq.s (\ref{scal_prop}) and (\ref{zasex}) exactly coincide with eq. (3.2) of
\cite{lw04}. This allows to relate the ``transition matrix elements'' reported
in eq. (C.6) of \cite{lw04} with the tension $T_0$ of the two $D0$ branes. 

It is clear from the above identifications that our derivation is fully
equivalent to the original one by L\"uscher and Weisz~\cite{lw04}. The only
advantage of our formulation is that it allows us to describe the result of this
mathematical transformation directly in the closed string formulation, as we
will see in the next section, sheding some more light on the string
interpretation of the transformation.
\FIGURE{
\psfrag{sigma}{\small $\sigma$}
\psfrag{tau}{\small $\tau$}
\includegraphics[width=4.2cm]{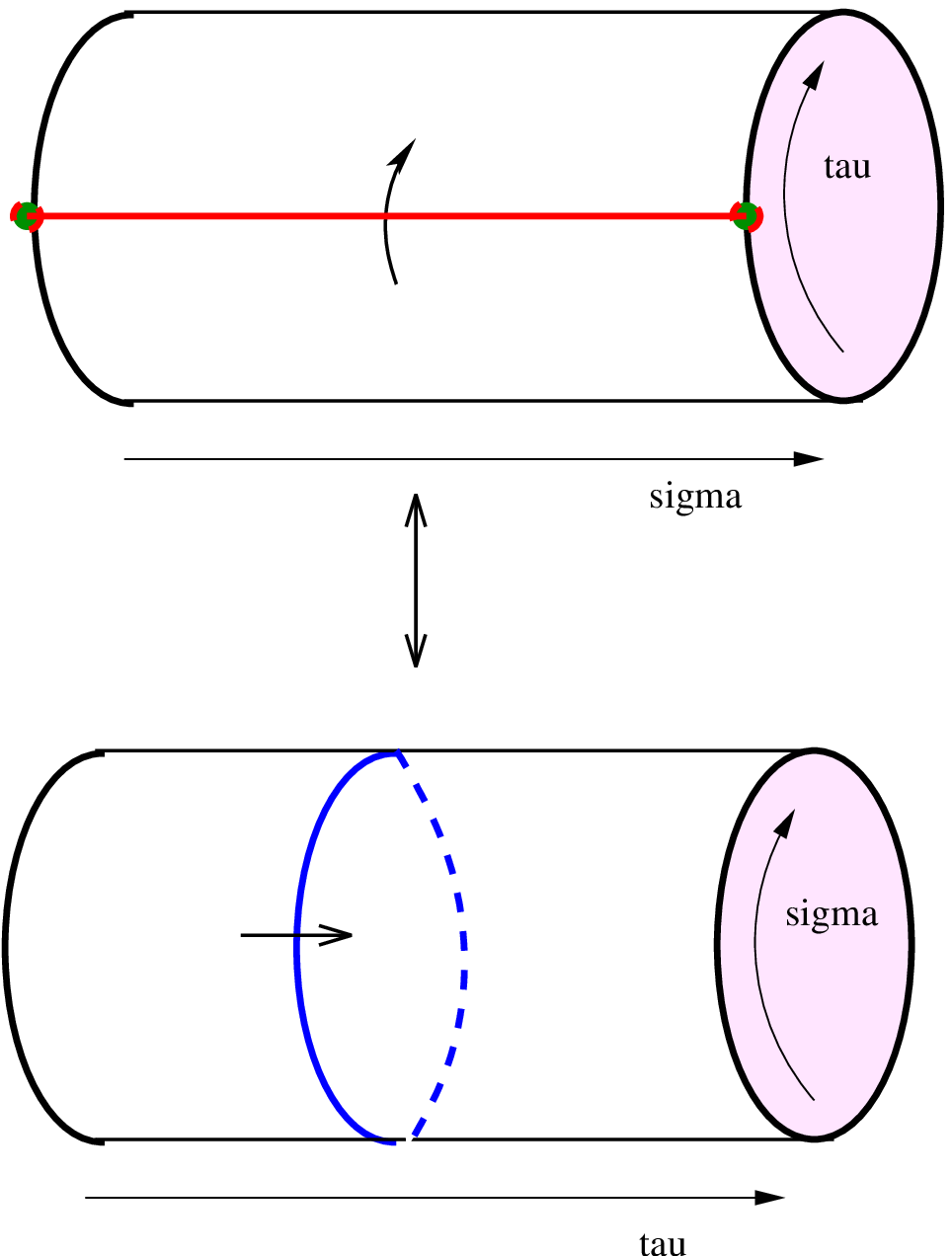}%
\caption{\label{fig:openclosed} 
\small The interchange $\sigma\leftrightarrow\tau$ maps the open string 1-loop
free energy (upper) to tree level propagation of a
closed string (lower).}
}

\subsection{Closed string interpretation}
\label{subsec:cl_bs}
Let us now re-derive the Nambu-Goto partition function in the closed string
channel. This requires the introduction of the notion of ``boundary state'',
which plays a major role in the following, and could be relevant for the stringy
treatment of other gauge geometries%
\footnote{See \cite{bs} for a list of references in the context of string and
superstring theory; ref. \cite{DiVecchia:1999rh} is a very useful review. See
\cite{Frau:1997mq} for a discussion of boundary states on compact target spaces,
which is relevant here for the $X^0$ direction. In ref. \cite{bs_CFT} we give
some basic references regarding boundary states in the context of CFT's. }. 

The modular transformation of the cylinder amplitude corresponds to the
interchange of the world-sheet coordinates $\sigma\leftrightarrow \tau$, so that
the loop of an open string gets re-interpreted as the tree level propagation of
a closed string between two \emph{boundary} states representing the two
D$0$-branes:
\beq
\label{tree_cl_D}
\mathcal{F} = \bra{B;\vec 0\,} 
\mathcal{D}
\ket{B;\vec R\,}~,
\eeq
where $\mathcal{D}$ is the closed string propagator, see below. An explicit
expression of the boundary states makes it possible to derive the closed-channel
expression \eq{mod3} of the free energy $\mathcal{F}$ entirely within the closed
string formalism.

The closed string fields $X^i(\tau,\sigma)$ in the spatial directions have the
standard mode expansion 
\beq
\label{Xic_mode}
X^i(\tau,\sigma) = \hat x^i + \alpha'\,\hat p^i\,\tau 
+ \ii \sqrt{\frac{\alpha'}{2}} \sum_{n\not=0}
\left(\frac{\alpha^i_n}{n}\ee^{-\ii n(\tau+\sigma)} +
\frac{\widetilde\alpha^i_n}{n}\ee^{-\ii n(\tau-\sigma)}
\right)~. 
\eeq
In the compact time direction, which has the periodicity \eq{comp_time}, the
zero-mode sector is modified:
\beq
\label{X0c_mode}
X^0(\tau,\sigma) = \hat x^0 + \alpha'\,\hat p^0\,\tau + \frac{\hat m
L}{2\pi}\,\sigma + \mbox{non-zero modes}~,
\eeq
where the momentum operator $\hat p^0$ has quantized eigenvalues $2\pi n/L$ and
$\hat m$ has integer eigenvalues $m$ corresponding to the \emph{winding number}
of the string in the time direction:
\beq
\label{wn}
X^0(\tau,\sigma+2\pi) = X^0(\tau,\sigma) + m L~.
\eeq 
The left- and right-moving oscillators $\alpha^M_n$ and $\widetilde\alpha^M_n$,
with $M=(0,i)$, satisfy the algebra
\beq
\label{osc}
\comm{\alpha^M_m}{\alpha^N_n} = m\,\delta_{m+n,0} \delta^{MN}
\eeq
(and the analogous one for the right-movers).
The left-moving and right-moving Virasoro generators $L_0$ and $\widetilde L_0$
are given by 
\beq
\label{L0Lt0}
\begin{aligned}
L_0 &= \frac{\alpha'}{4} \sum_i (\hat p^i)^2 + 
\frac{\alpha'}{4} \left(\hat p^0  + \frac{\hat m
L}{2\pi\alpha'}\right)^2 
+ \sum_{n=1}^\infty \alpha_{-n}\cdot \alpha_n -\frac{d}{24}~,
\\
\widetilde L_0 &= 
\frac{\alpha'}{4} \sum_i (\hat p^i)^2 + 
\frac{\alpha'}{4} \left(\hat p^0  - \frac{\hat m
L}{2\pi\alpha'}\right)^2 
+ \sum_{n=1}^\infty \widetilde\alpha_{-n}\cdot
\widetilde\alpha_n - \frac{d}{24}~,
\end{aligned}
\eeq
where we included the normal ordering constants. A standard form of the closed
string propagator is given by
\beq
\label{csp}
\frac{\alpha'}{4\pi} 
\int \frac{d^2 z}{|z|^2} z^{L_0+L_0^{\mathrm{gh.}}} \bar z^{\widetilde L_0 +
\widetilde L_0^{\mathrm{gh.}}}~,
\eeq
where $L_0^{\mathrm{gh.}}$ and  $\widetilde
L_0^{\mathrm{gh.}}$ are the Virasoro generators for the ghost system. 

The \emph{boundary state} $\ket{B;\vec R}$ appearing in \eq{tree_cl_D}
represents in the closed string language a D$0$-brane located in its transverse
directions at $\vec R$. It is a state in the Hilbert space of the closed string
which inserts a boundary in the world-sheet at $\tau=0$ and enforces the closed
string counterparts of the open string boundary conditions eq.s
(\ref{nbcp},\ref{dbcp}), Neumann along the time, Dirichlet along the spatial
directions:
\beq
\label{bs_cond}
\left.\partial_\tau X^0(\sigma,\tau)\right|_{\tau=0} \ket{B;\vec R} = 0~,
\hskip 0.8cm 
\left.\left(X^i(\sigma,\tau) - R^i \right) \right|_{\tau=0} \ket{B;\vec
R} =0 ~.
\eeq    
In terms of the mode expansions \eq{Xic_mode} and \eq{X0c_mode}, these
conditions become
\beq
\label{ncl}
\left(\alpha^0_n + \widetilde \alpha^0_{-n}\right) \ket{B;\vec R\,} = 0~,
\hskip 0.8cm
\hat n \ket{B;\vec R\,} = 0
\eeq
in the time directions, and
\beq
\label{dcl}
\left(\alpha^i_n - \widetilde \alpha^i_{-n}\right) \ket{B;\vec R\,}  = 0~,
\hskip 0.8cm
\left(\hat x^i - r^i\right)\ket{B;\vec R\,}  = 0~.
\eeq
in the spatial ones. It follows from these conditions that the boundary state
also identifies the left- and right-moving Virasoro generators:
\beq
\label{L0=L0t}
\left(L_0 - \widetilde L_0\right)\ket{B;\vec R\,} = 0~.
\eeq
Requiring the BRST invariance of the boundary state implies that it also has a
component in the ghost sector; for the details  we refer, for instance, to
\cite{DiVecchia:1999rh}. Let us just notice that the boundary state identifies
also the left- and right-moving ghost Virasoro generator
$L_0^{\mathrm{gh}},\widetilde L_0^{\mathrm{gh}}$, analogously to \eq{L0=L0t}.

The solution to these conditions has the form
\beq
\label{bsc}
\ket{B;\vec R\,} = \mathcal{N}_0 \,
\ket{B^{(0)};\vec R\,}
\ket{B^{\mathrm{n.z.}}}\ket{B^{\mathrm{gh.}}}~,
\eeq
where $\mathcal{N}_0$ is a normalization to be fixed. The non-zero-mode part of
the boundary state reads explicitly
\beq
\label{bsnzm}
\ket{B^{\mathrm{n.z.}}} =
\exp\left\{-\sum_{n=1}^\infty \frac 1n 
\left(\alpha^0_{-n} \widetilde \alpha^0_{-n} - 
\sum_i \alpha^i_{-n}\cdot \widetilde \alpha^i_{-n}
\right)\right\} 
\ket{0,\widetilde 0}~.
\eeq
Since the Neumann conditions \eq{ncl} leave the winding $m$ undetermined, the
zero-mode part of the boundary state decomposes as $\ket{B^{(0)};\vec R\,} =
\sum_m \ket{B^{(0)};\vec R\,;m}$, the $m$-th component belonging to the
sub-sector of the Hilbert space that describes strings wrapped $m$ times around
the time circle. We have
\beq
\label{bszm}
\ket{B^{(0)};\vec R\,;m}
= \delta(\hat{\vec x} - \vec
R) \ket{n=0,m; \vec p=0}
= 
\int\! \frac{d^{d-1}p}{(2\pi)^{d-1}}\ee^{-\ii \vec p\cdot \vec R}
\ket{n=0,m;\vec p}~,
\eeq
where in the second step  we made explicit the $\delta$-function along the
Dirichlet directions and introduced momentum eigenstates%
\footnote{These states we normalize to $\braket{\vec k}{\vec p} =
(2\pi)^{d-1}\delta^{d-1}(\vec k - \vec p)$. The quantized momentum and winding
$\ket n$ and $\ket m$ in the time direction are instead simply normalized to the
Kronecker $\delta$.}
$\ket{\vec p}$. 

Taking into account the identification of the left- and right-moving Virasoro
generators on the boundary states, the amplitude \eq{tree_cl_D} becomes,
introducing $|z| = \exp(-\pi s)$,
\beq
\label{tree_cl2}
\mathcal{F} = \frac{\pi\alpha'}{2} \int_0^\infty ds\, \bra{B;\vec 0\,} 
\ee^{-2\pi s(L_0+L_0^{\mathrm{gh.}})} 
\ket{B;\vec R\,}~. 
\eeq
The matrix element for the non-zero mode sector of the above expression can be
easily computed using the oscillator algebra \eq{osc} and the properties of
coherent-like states such as the ones appearing in \eq{bsnzm}, see
\cite{DiVecchia:1999rh}. The result is
\beq
\label{me_nzm}
\bra{B^{\mathrm{n.z.}}}  
\ee^{-2\pi s \left(\sum_{k=1}^\infty \alpha_{-k}\cdot \alpha_k -
\frac{d}{24}\right) } 
\ket{B^{\mathrm{n.z.}}} 
= \ee^{\frac{\pi d s}{12}}
\prod_{n=1}^\infty \left(\frac{1}{1 - \ee^{-2\pi n s}}\right)^d
= \left(\frac{1}{\eta(\ii s)}\right)^d~.
\eeq
The matrix element in the ghost sector effectively cancels out the 
contributions of the bosonic non-zero modes of two directions:
\beq
\label{me_gh}
\bra{B^{\mathrm{gh.}}}
\ee^{-2\pi s L_0^{\mathrm{gh.}}}
\ket{B^{\mathrm{gh.}}}
= \eta^2(\ii s)~.
\eeq
The matrix element in the zero-mode sector, using \eq{bszm} and the 0-mode part
of $L_0$ as given in \eq{L0Lt0}, becomes, for each winding $m$,
\beq
\label{me_zm}
\begin{aligned}
& 
\int\! \frac{d^{d-1}q}{(2\pi)^{d-1}}
\int\! 
\frac{d^{d-1}p}{(2\pi)^{d-1}}
\bra{n\!\!=\!\!0,m;\vec q}
\ee^{-\frac{\pi\alpha' s}{2} 
\left(\sum_i (\hat p^i)^2  + \left(\hat p^0 + \frac{\hat m
L}{2\pi\alpha'}\right)^2\right) -\ii \vec p\cdot \vec R}
\ket{n\!\!=\!\!0,m;\vec p}
\\
& =  (2\pi^2 \alpha')^{-\frac{d-1}{2}} \ee^{-\frac{m^2 L^2
s}{8\pi\alpha'} -\frac{r^2}{2\pi\alpha's}}~,
\end{aligned}
\eeq 
where in the last step we used the orthogonality of the momentum and winding
eigenstates, and performed the remaining Gaussian integration. 

The amplitude $\mathcal{F} = \sum_m \mathcal{F}^{(m)}$ receives contributions
from all winding sectors and, collecting all the ingredients, we find
\beq
\label{zc}
\mathcal{F}^{(m)} = \mathcal{N}^2_0\,(2\pi^2\alpha')^{\frac{1-d}{2}}
\frac{\pi\alpha'}{2}
\int_0^\infty \frac{ds}{s} s^{\frac{3-d}{2}}
\ee^{-\frac{m^2 L^2\,s}{8\pi\alpha'}-
\frac{R^2}{2\pi\alpha' s}}
\left(\frac{1}{\eta(\ii s)}\right)^{d-2}~.
\eeq
The boundary state normalization has to be chosen so as to agree with the
modular transformation, \eq{mod3}, of the open channel cylinder trace. This
fixes
\beq
\label{Np}
\mathcal{N}_0 = \frac{T_0}{2}\sqrt{L}~,
\eeq
where the tension $T_p$ of the bosonic D$p$-brane in the non-compact theory
was already given in \eq{Tp}.

\section{Conclusions.}
\label{sec:conclusions}
In this paper we have shown how to derive the Nambu-Goto effective string from a
covariant quantization of the bosonic string in $d$ dimensions, which is
tantamount to the Polyakov formulation if one neglects the Liouville field.
     
As we noticed in the introduction, for large enough values of $R$ and $L$ the
Nambu-Goto string is in very good agreement with the results of Montecarlo
simulations for various gauge theories and in various dimensions, see for
instance Fig. \ref{fig80new} and  \ref{fig:ising} (taken with small changes from
\cite{chp05}) referring to the $\mathbb{Z}_2$ gauge theory in (2+1) dimensions. 
\FIGURE{
\includegraphics[width=12cm]{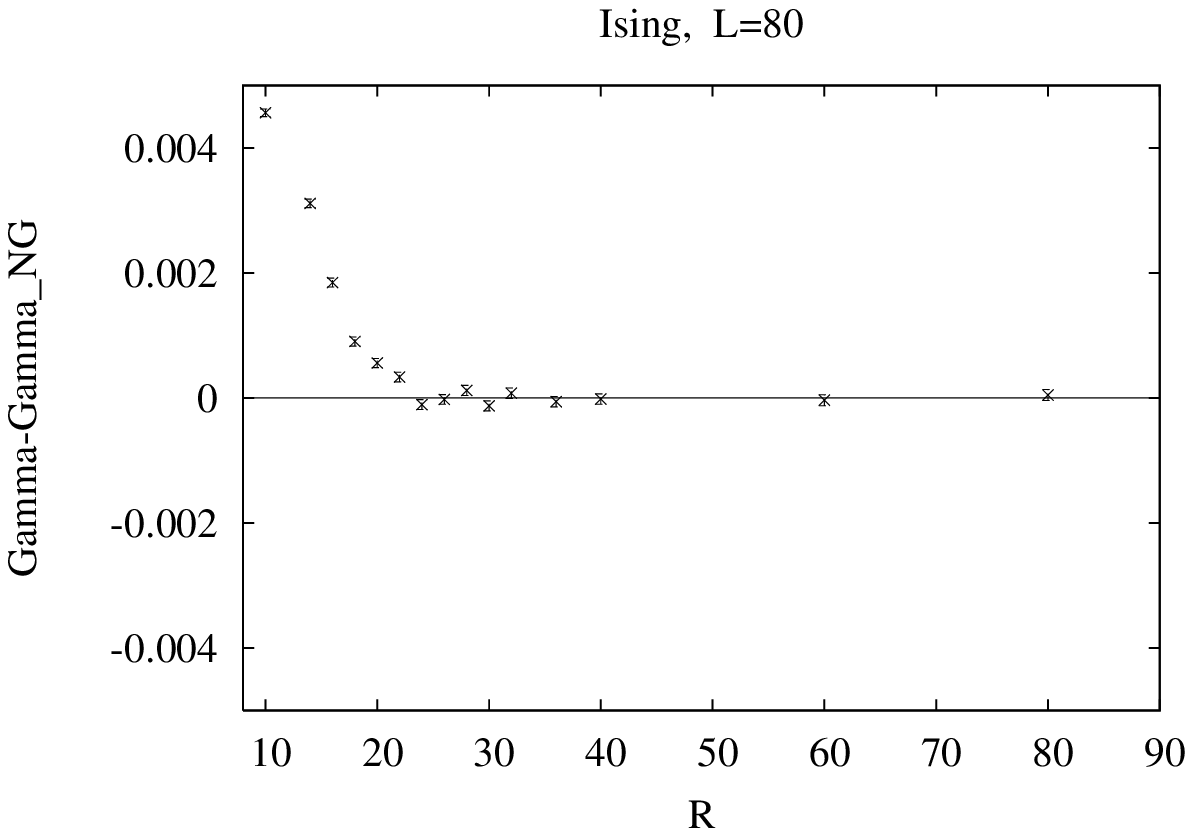}
\caption{\small Montecarlo results for the Polyakov loop correlators in the
(2+1) dimensional gauge Ising model. The data are taken at a fixed value of the
lattice in the time direction: $L=80$ (which corresponds to a very low
temperature $T=T_c/10$) and a varying size of the interquark distance
($10<R<80$). In the figure is plotted the deviation of $\Gamma$ (the ratio
$G(R+1)/G(R)$ of two correlators shifted by one lattice spacing, see
\cite{chp05} for details) with respect to the Nambu-Goto string expectation
$\Gamma_{NG}$ (which with this definition of observables corresponds to the
straight line at zero). Notice the remarkable agreement in the range $24<R<80$,
which is not the result of a fitting procedure: in the comparison reported in
the figure there is no free parameter (data taken from ref. \cite{chp04} and
\cite{chp05}).}
\label{fig80new}
}
\FIGURE{
\includegraphics[width=12cm]{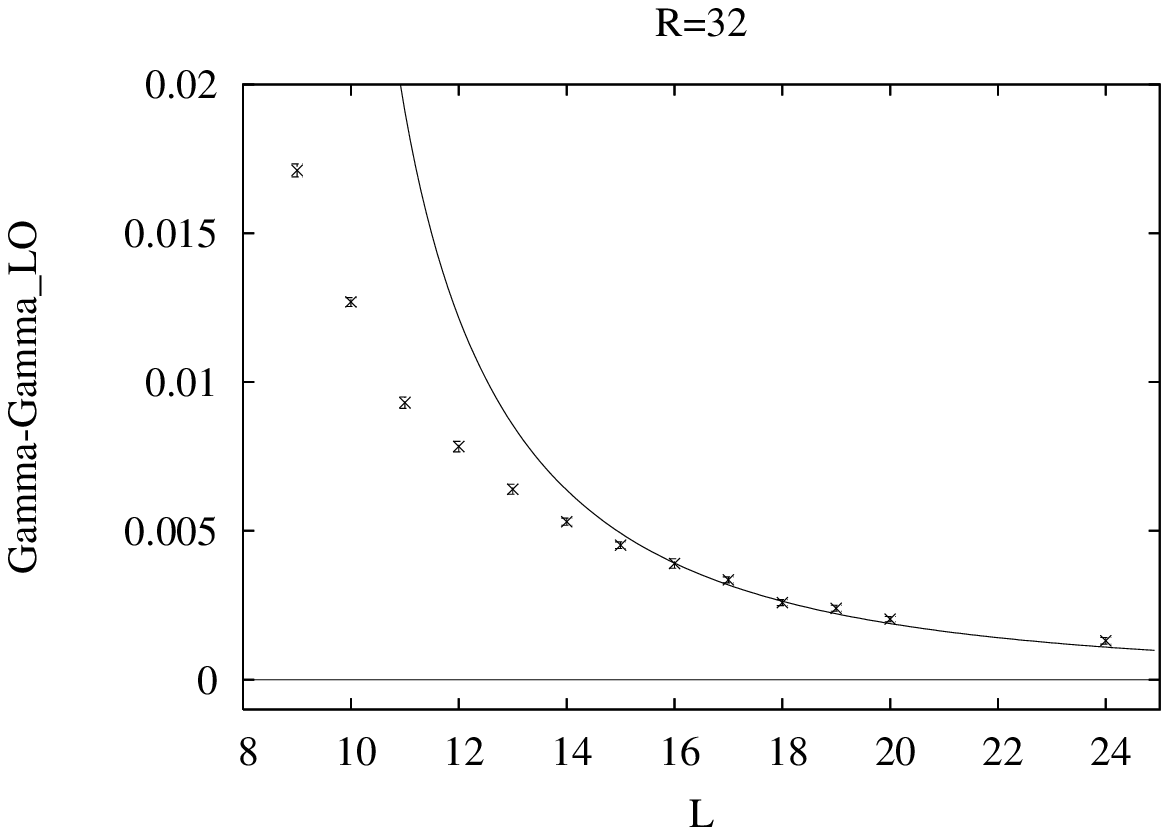}
\caption{\small Montecarlo results for the Polyakov loop correlators in the
(2+1) dimensional gauge Ising model. The data are taken at a fixed value  $R=32$
of the interquark distance and a varying size ($8<L<24$) of the lattice in the
time directions. In the figure is plotted the deviation of $\Gamma$ (defined as
in the previous figure) with respect to the asymptotic free string expectation
$\Gamma_{LO}$ (which with this definition of observables corresponds to the
straight line at zero). The curve is the Nambu-Goto prediction for this
observable. Notice the remarkable agreement in the range $16<L<24$, which as
for the previous figure is not the result of a fitting procedure: in the
comparison reported in the figure there is no free parameter (taken from Fig. 3
of ref.\cite{chp05}).}
\label{fig:ising}
}
Work is in progress \cite{chp05_new} to quantify this agreement also at the
level of the string spectrum, and the preliminary results are very favourable to
the Nambu-Goto effective model. This suggests that indeed at large distance one
can neglect the Liouville mode and still correctly describe the fluctuations of
the flux tube which joins together the quark--anti-quark pair by standard
bosonic string theory. As it has been observed by several 
groups~\cite{chp04,chp05,jkm03,jkm04}, and as it is
clearly visible in Fig.s \ref{fig80new} and \ref{fig:ising}, at shorter
distances the Montecarlo data show a drastic deviation from the Nambu-Goto
predictions. It would be interesting to understand if this signals the breakdown
of the string picture itself, or if it is possible to describe these deviations
in a string framework. In this respect, one would like to understand whether
also the \emph{deviations} from the Nambu-Goto behaviour follow a universal
pattern. Preliminary results indicate that this is indeed the case for the Ising
model and  the SU(2) models in (2+1) dimensions, see \cite{cpr04}. A similar,
but not completely coincident behaviour has been observed for the SU(3) model
\cite{cpr04,chp05}. Further numerical studies, auspiciously in a wider range of
models, are needed to completely clarify this point. Nevertheless, we can ask
ourselves which stringy mechanism could possibly be responsible for these
deviations. A r\^ole could be played by contributions from
higher genus surfaces bounded by the Polyakov
loops; our formalism could help to test this possibility. 

A second interesting possibility is that at short distance one is not any more
allowed to neglect the Liouville mode in the analysis, and should resort to the
full Polyakov formulation. To this purpose, the preliminary numerical results of
\cite{chp05_new} seem to suggest that the modification for shorter scales
consists basically in a shift of the spectrum, in a way compatible with the
effect of an extra degree of freedom.  

At first sight, including another degree of freedom seems to contradict
the expected behaviour of the flux tube, whose relevant d.o.f. should be the
$d-2$ transverse oscillations only. For this reason, after the discussions of
the eighties and early nineties, no standard consistent quantization of the
bosonic string seemed suited to describe the flux tube dynamics at all scales
$R$: in $d<26$, all exhibit unwanted features%
\footnote{In a non-covariant quantization of the free bosonic string Lorentz
symmetry is broken, while in covariant quantization extra longitudinal modes
appear.}, including the Polyakov string, which has an extra dimension
parametrized by the Liouville field. Some interesting alternative proposals were
made, such as, for instance, the effective string theory of Polchinski and
Strominger \cite{Polchinski:1991ax}, but a clear picture and a single candidate
did not emerge. 

More recently, in particular through the work of Polyakov \cite{Pol_cave} and
the progresses made in supersymmetric Yang-Mills theory through the AdS/CFT
correspondence \cite{maldacena}, a different conceptual picture has been
developed. The extra dimension parametrized by the Liouville field should
represent the renormalization scale of the quantum gauge theory, rather than
another space-time direction, thus evading the objection mentioned above to the
relevance of the Polyakov model for the QCD string. 

This paper aims not to tackle the difficult analysis of the relation between the
full-fledged Polyakov formulation and gauge theories. Yet, having retrieved the
large-$R$ behaviour described by the Nambu-Goto effective model (and favoured by
the numerical simulations) in a first-order formulation \emph{\`a la} Polyakov,
but neglecting the Liouville field, might prove useful in discussing the
modifications due to its inclusion.
 
\acknowledgments{We thank G. Delfino, F. Gliozzi, M. Hasenbusch, A. Lerda, M.
Panero, I. Pesando, R. Russo and S. Sciuto for very useful discussions.}

\end{document}